\documentclass[prx, preprintnumbers, amsmath, amssymb, superscriptaddress, twocolumn]{revtex4-1}
\usepackage{graphicx, color, dcolumn, natbib, bm, amssymb, amsmath, soul}
\usepackage[table,  dvipsnames]{xcolor}
\graphicspath{{fig/}}
\usepackage[final=true]{hyperref}
\usepackage[normalem]{ulem}

\begin{document}
\title{Robust metastable skyrmions with tunable size in the chiral magnet FePtMo$_3$N}
\author{A. S. Sukhanov}
\affiliation{Max Planck Institute for Chemical Physics of Solids, D-01187 Dresden, Germany}
\affiliation{Institut f{\"u}r Festk{\"o}rper- und Materialphysik, Technische Universit{\"a}t Dresden, D-01069 Dresden, Germany}
\author{A. Heinemann}
\affiliation{German Engineering Materials Science Centre (GEMS) at Heinz Maier-Leibnitz Zentrum (MLZ), Helmholtz-Zentrum Geesthacht GmbH, D-85747 Garching, Germany}
\author{L. Kautzsch}
\affiliation{Materials Research Laboratory, University of California, Santa Barbara, California 93106, USA}
\affiliation{Max Planck Institute for Chemical Physics of Solids, D-01187 Dresden, Germany}
\author{J. D. Bocarsly}
\affiliation{Materials Research Laboratory, University of California, Santa Barbara, California 93106, USA}
\affiliation{Materials Department, University of California, Santa Barbara, California 93106, USA}
\author{S. D. Wilson}
\affiliation{Materials Research Laboratory, University of California, Santa Barbara, California 93106, USA}
\affiliation{Materials Department, University of California, Santa Barbara, California 93106, USA}
\author{C. Felser}
\affiliation{Max Planck Institute for Chemical Physics of Solids, D-01187 Dresden, Germany}
\author{D. S. Inosov}
\affiliation{Institut f{\"u}r Festk{\"o}rper- und Materialphysik, Technische Universit{\"a}t Dresden, D-01069 Dresden, Germany}
\begin{abstract}

Synthesis of new materials that can host magnetic skyrmions and their thorough experimental and theoretical characterization are essential for future technological applications. The $\beta$-Mn-type compound FePtMo$_3$N is one such novel material that belongs to the chiral space group $P4_132$, where the antisymmetric Dzyaloshinkii-Moriya interaction is allowed due to the absence of inversion symmetry. We report the results of small-angle neutron scattering (SANS) measurements of FePtMo$_3$N and demonstrate that its magnetic ground state is a long-period spin helix with a Curie temperature of 222~K. The magnetic field-induced redistribution of the SANS intensity showed that the helical structure transforms to a lattice of skyrmions at $\sim$13~mT at temperatures just below $T_{\text C}$. Our key observation is that the skyrmion state in FePtMo$_3$N is robust against field cooling down to the lowest temperatures. Moreover, once the metastable state is prepared by field cooling, the skyrmion lattice exists even in zero field. Furthermore, we  show that the skyrmion size in FePtMo$_3$N exhibits high sensitivity to the sample temperature and can be continuously tuned between 120 and 210~nm. This offers new prospects in the control of topological properties of chiral magnets.

\end{abstract}
 
\maketitle

Despite the active search for novel materials that host topologically-nontrivial whirls of magnetic moments, known as magnetic skyrmions, the number of discovered compounds remains low. The chiral magnet MnSi, the first compound where the skyrmion lattice (SkL) phase was found, belongs to the noncentrosymmetric space group $P2_13$~\cite{Muehlbauer}. As was later recognized, more compounds with the $P2_13$ space group demonstrate skyrmions, such as FeGe~\cite{Yu_2,Wilhelm,Moskvin}, Fe$_{1-x}$Co$_x$Si~\cite{Muenzer,Grigoriev07,Yu},  Mn$_{1-x}$Fe$_x$Si~\cite{Pfleiderer,Bauer,Grigoriev09}, and Cu$_2$OSeO$_3$~\cite{Seki12,Adams}. All these materials are characterized by a generic phase diagram that is well understood in terms of four competing magnetic interactions: the Heisenberg exchange interaction, the Dzyaloshinskii-Moriya interaction (DMI), a weak cubic magnetocrystalline anisotropy, and the Zeeman energy~\cite{Bak,Grigoriev15,Roessler,Bogdanov}. The competition between the ferromagnetic exchange $J$ and the DMI constant $D$ results in a spin-spiral ground state with the spiral period $\lambda \propto J/D$. The cubic anisotropy, in turn, determines the orientation of the helical propagation vector $\textbf{q}$ in the absence of applied magnetic field. Upon application of the magnetic field, the helical magnetic structure either transforms into a conical spiral or  forms the SkL. The delicate energy balance between the two field-induced states is heavily influenced by thermal fluctuations~\cite{Muehlbauer,Laliena,Buhrandt}.

The thermal fluctuations lower the energy of the skyrmion state, but it is only in the narrow region below $T_{\text C}$ where the SkL becomes energetically favored over the conical spiral in the cubic chiral magnets. In many real materials, the SkL phase pocket extends in temperature for only a few percent of the magnetic ordering temperature, which significantly limits potential applications of these materials in future technologies~\cite{Muehlbauer,Yu_2,Wilhelm,Moskvin,Muenzer,Grigoriev07,Yu,
Pfleiderer,Bauer,Grigoriev09,Seki12,Adams}. In regard to this problem, another class of the chiral magnets, the $\beta$-Mn-type Co-Zn-Mn compounds, stand out as the materials where a robust metastable SkL is observed in a very wide temperature range~\cite{Tokunaga,Karube4,Ukleev}. To achieve this, a field-cooling (FC) process through the equilibrium skyrmion phase can be utilized, as was demonstrated for Co$_8$Zn$_8$Mn$_4$~\cite{Karube} and Co$_9$Zn$_9$Mn$_2$~\cite{Karube3}. Similarly to the $P2_13$ compounds, the DMI in Co-Zn-Mn compounds is caused by the absence of the inversion symmetry in the $P4_132$ or $P4_332$ space group (depends on the handedness) they belong to. Moreover, not only can the SkL state be quenched to much lower temperatures below $T_{\text C}$ in Co$_9$Zn$_9$Mn$_2$, but also it remains robust when the magnetic field is removed~\cite{Karube3}. It was suggested that the site occupancy disorder inherent to the Co-Zn-Mn structure might be essential for the observed metastability~\cite{Bocarsly,Nakajima}.

The materials that exhibit robust metastable skyrmions in a wide temperature and magnetic-field range, including zero field, open up ways for a broader range of applications~\cite{Fert}. Thus, it is very important to explore if more compounds realizing such properties can be found. Recently, the molybdenum nitrides with the $\beta$-manganese structure and the general formula $A_2$Mo$_3$N (where $A$ are transition metals) were discussed as new candidate skyrmion-hosting materials~\cite{Li}. Particularly, Lorentz transmission electron microscopy (LTEM) measurements in a thin polycrystalline plate of FeCo$_{0.5}$Rh$_{0.5}$Mo$_3$N confirmed the presence of skyrmion clusters in the vicinity of grain boundaries~\cite{Li}.   Surprisingly, the magnetic susceptibility of a bulk sample, analyzed in the same study, did not demonstrate any signatures of the SkL state. This suggested that the results of the LTEM observations may be linked to the confined geometry of a thin sample plate ($\sim$100~nm thickness).

In this paper, we focus on another representative of the chiral magnets with the $\beta$-Mn-structure -- FePtMo$_3$N (space group $P4_132$) -- and demonstrate that its magnetic ground state is a long-wavelength spin spiral. Small-angle neutron scattering (SANS), as a reciprocal-space imaging technique, allowed us to observe the field-induced SkL phase, which was previously suggested by dc and ac magnetic measurements (the A phase)~\cite{Kautzsch}. Further, we show that field-cooling through the stable skyrmion pocket results in the formation of robust metastable skyrmion lattice that persists to low temperature and zero field.


The SANS measurements were conducted at the instrument SANS-1 at the FRM-II (Garching, Germany).
A polycrystalline sample used in the present measurements was synthesized as described in Ref.~\cite{Kautzsch}.

\begin{figure}[t]
\includegraphics[width=0.99\linewidth]{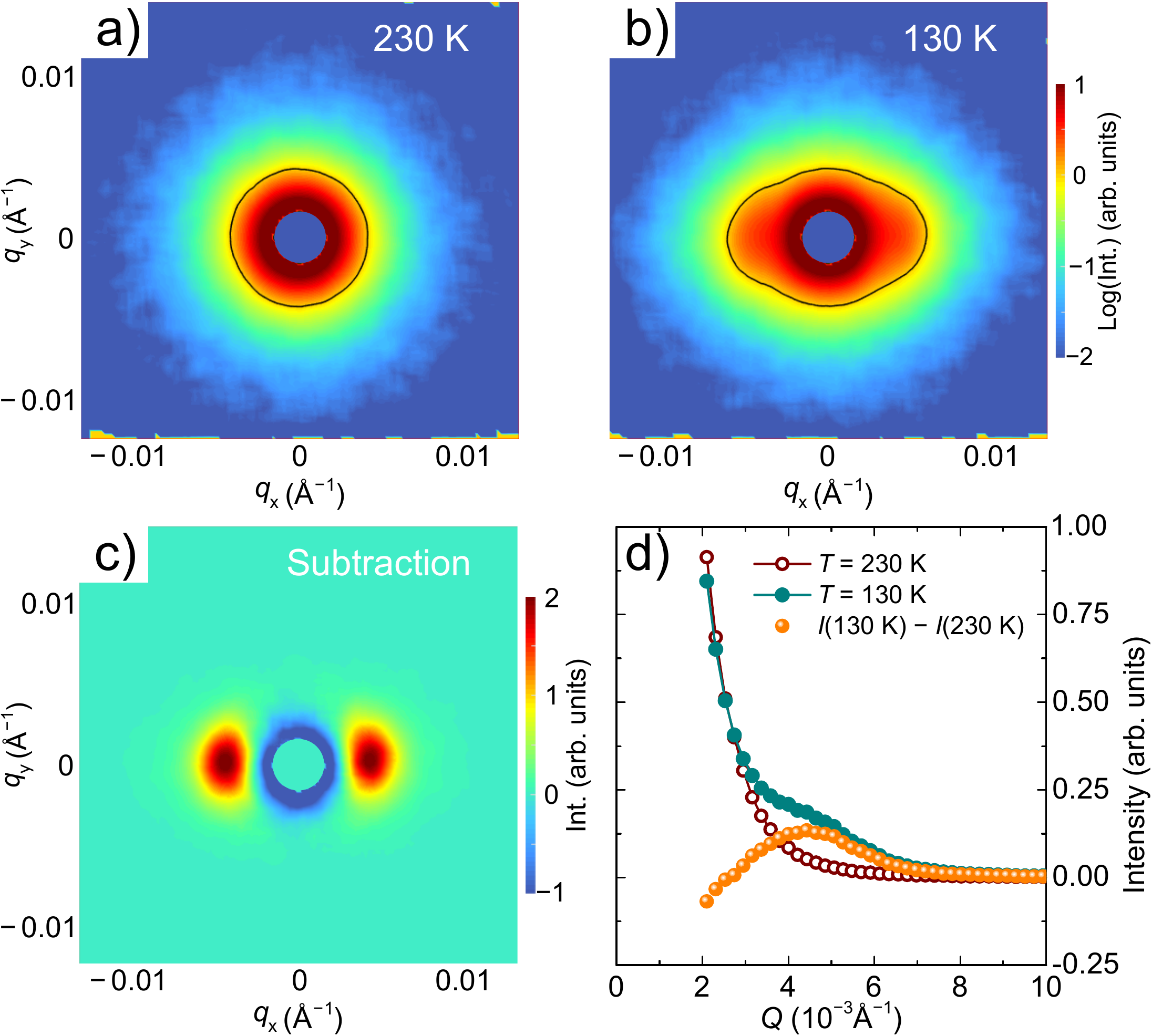}\vspace{3pt}
        \caption{(color online). SANS scattering from the helical structure of FePtMo$_3$N. (a) A SANS pattern collected above the magnetic ordering temperature at $T = 230$~K. (b) A SANS pattern at $T = 130$~K, well below $T_{\text C}$. (c) The difference between the two patterns, $I$(130~K)$-I$(230~K). (d) The scattering intensity as a function of momentum transfer $Q$. The solid lines are a guide to the eye.}
        \label{ris:fig1}
\end{figure}

Figure~\ref{ris:fig1}(a) shows a SANS pattern collected at the sample temperature of 230~K, which is just above $T_{\text C}$. The scattering pattern shows an isotropic distribution of intensity $I$, which rapidly decays with increasing momentum $Q$ according to the $I \propto Q^{-4}$ law. Such scattering is known as the Porod scattering and occurs due to a nuclear contrast at the surface of individual crystallites within the sample. Thus, it is considered as a background. The magnetic neutron scattering is clearly evidenced in Fig.~\ref{ris:fig1}(b), which demonstrates a typical SANS pattern collected below $T_{\text C}$. Unlike the isotropic pattern at $T = 230$~K, the scattering at $T = 130$~K features significant elliptical anisotropy. To highlight the contribution from the helical magnetic texture, we subtracted the intensity distribution of $I(\textbf{Q}, 230~{\text K})$ from the pattern $I(\textbf{Q}, 130~{\text K})$. Figure~\ref{ris:fig1}(c) depicts the resulting magnetic contribution. As can be seen, the spin spiral state is characterized by a pair of Bragg peaks located at the momentum $Q = q \approx 0.0045$~\AA$^{-1}$, where $q$ is the helical propagation vector. Typically, a polycrystalline helimagnet shows an isotropic ring of intensity at $|\textbf{Q}| = q$, as individual grains all have different orientations of the propagation vector. Here, cooling under a small remanent magnetic field of $\sim$2~mT is evidently sufficient to cause all of the helices to rotate along the field direction, resulting in a characteristic two-fold symmetry of the pattern.


\begin{figure}[t]
\includegraphics[width=0.99\linewidth]{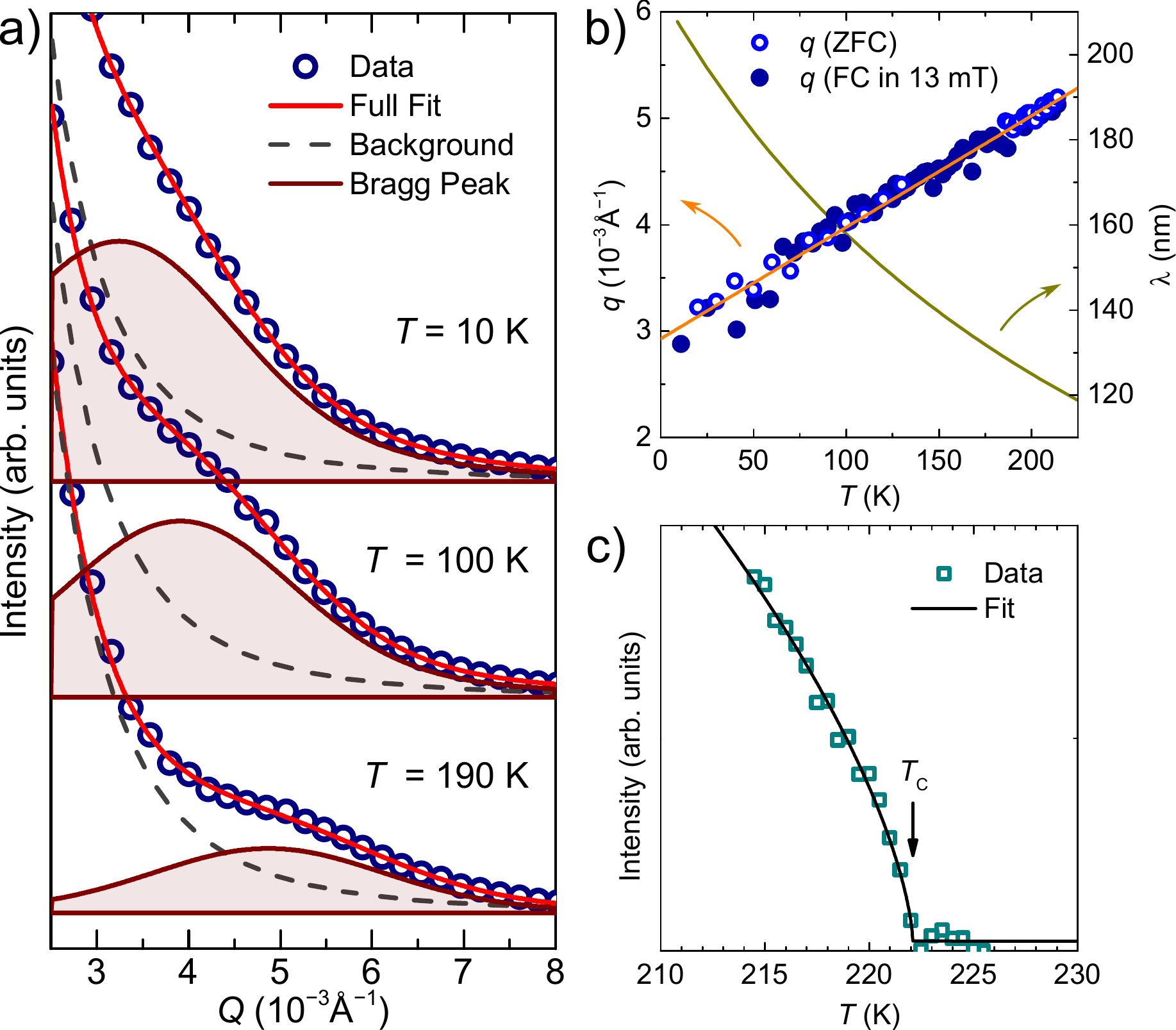}\vspace{3pt}
        \caption{(color online). Temperature variation of the propagation vector. (a) SANS intensity profiles at different temperatures below $T_{\text C}$. The open circles are experimental data, the red solid and the black dotted lines are the full fitting curve and the background contribution, respectively. The shaded area is the Bragg peak contribution. (b) The propagation vector as a function of temperature, the solid line is a linear fit. (c) The intensity of the helical Bragg peak in the vicinity of $T_{\text C}$. The solid line is a fit by the critical exponent function.}
        \label{ris:fig2}
\end{figure}

\begin{figure}[t]
\includegraphics[width=0.99\linewidth]{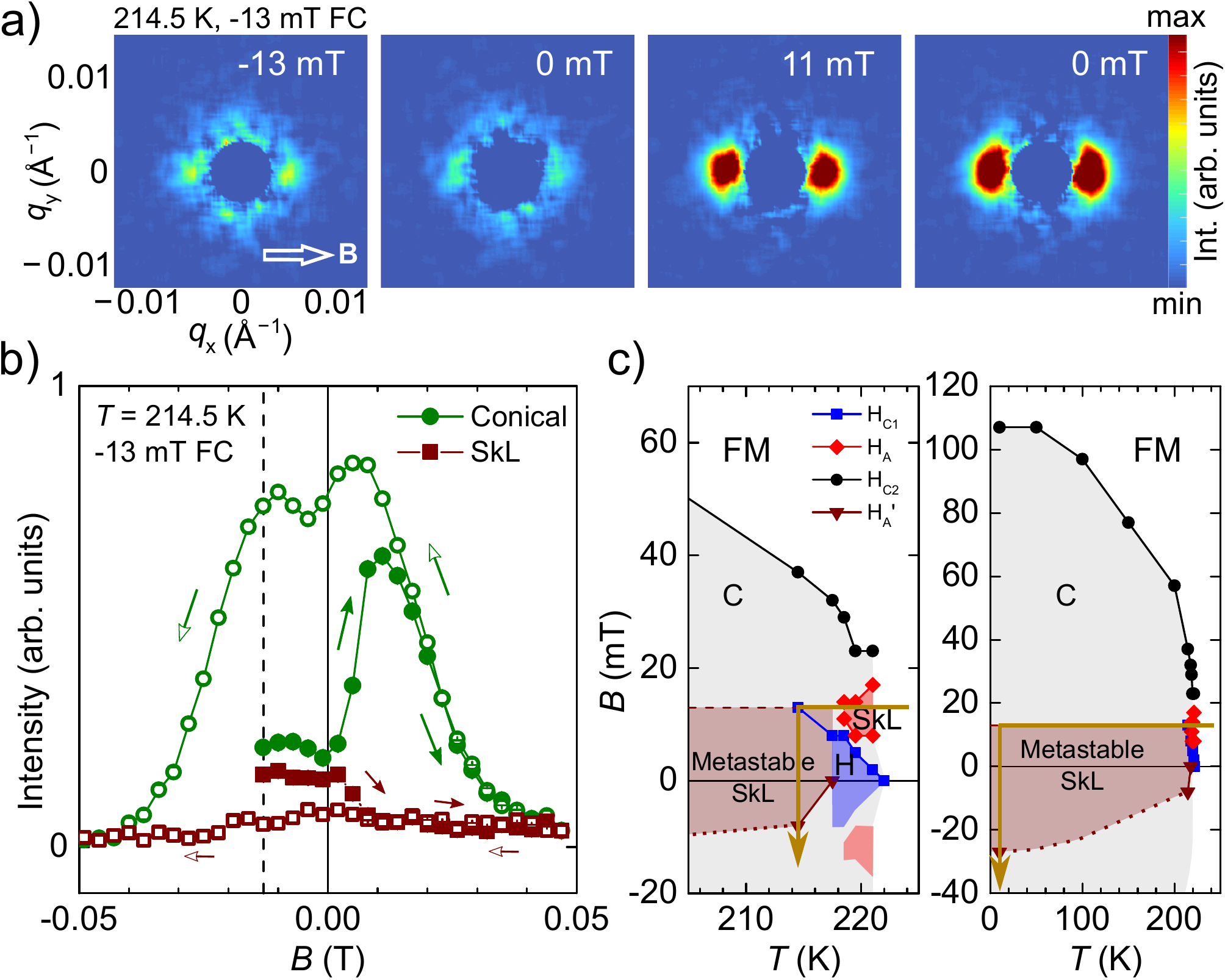}\vspace{3pt}
        \caption{(color online). The metastability of the SkL phase at zero field. (a) A series of SANS patterns (after subtraction of the paramagnetic scattering) recorded at $214.5$~K at different magnetic fields after $-13$~mT FC from $T > T_{\text C}$. (b) The intensities of the SkL and the conical-phase Bragg peaks as a function of applied field. The closed symbols show the data collected in the first field sweep, the open symbols denote the second sweep. (c) The tentative phase diagram with the phase boundaries determined by SANS. The discussed field history is marked by the arrow.}
        \label{ris:fig3}
\end{figure}

\begin{figure}[t]
\includegraphics[width=0.99\linewidth]{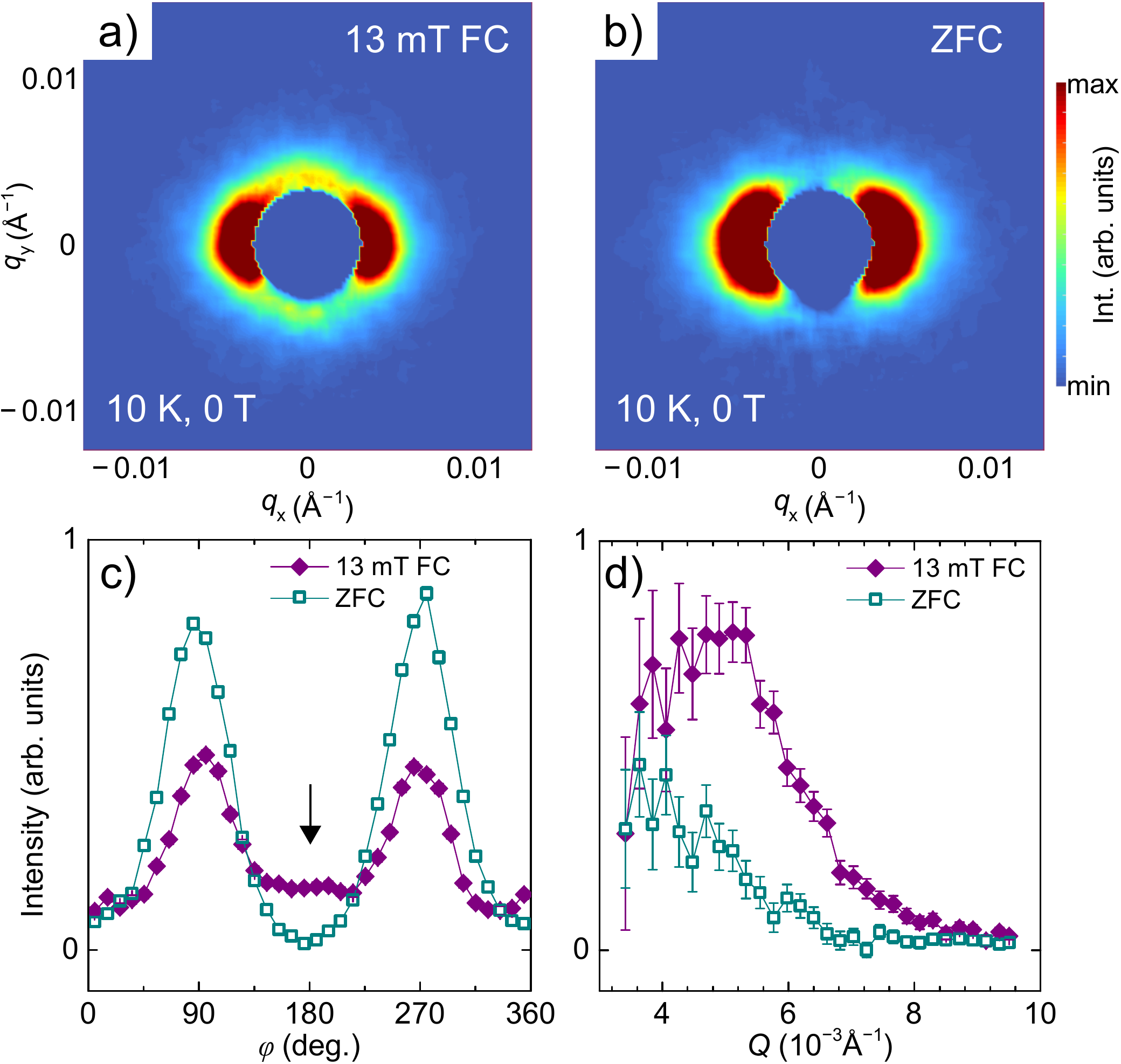}\vspace{3pt}
        \caption{(color online). SANS scattering (after subtraction of the $T > T_{\text C}$ scattering) at $T = 10$~K and $B = 0$~mT after FC and ZFC procedures. (a) A SANS pattern recorded at zero field after FC in $-13$~mT. (b) A SANS pattern at the same final conditions but after ZFC. (c) The intensity profiles extracted from two patterns as a function of the in-plane angle $\phi$. (d) The $Q$-dependence of the scattering intensities after ZFC and FC at $\phi = 0^{\circ}$, $180^{\circ}$.}
        \label{ris:fig4}
\end{figure}

The intensity profiles $I(Q_x)$ were extracted from the SANS maps and plotted in Fig.~\ref{ris:fig1}(d). To plot the profiles, we averaged the intensity over two opposite 15$^{\circ}$ sectors centred at the $Q_x$ axis. After a detailed comparison of the profiles in the ordered state and in the paramagnetic phase, one can notice that the magnetic fluctuations above $T_{\text C}$ cause some observable quasielastic scattering in the vicinity of $Q = 0$. Therefore, when $I(Q, 230~{\text K})$ is subtracted from the $T = 130$~K pattern, an isotropic negative contribution appears at $Q < 0.002$~\AA$^{-1}$, which slightly affects the shape of the magnetic Bragg peaks originating from the helical structure (also evident in Fig.~\ref{ris:fig1}(c)). To avoid this, we found it more convenient to analyze the intensity profiles without the background subtraction. Instead, we take the background into account by a model function in every profile.

After confirming that FePtMo$_3$N is a helimagnet and determining the scale of the spiral period, we turn to the temperature variation of the propagation vector. Figure~\ref{ris:fig2}(a) shows examples of the intensity profiles at different temperatures spanning a wide range, $T = 10$, 100, and 190~K. The profiles were fitted by a model function that takes into account a background (the Porod scattering) and the magnetic Bragg peak (described by a Gaussian function).
Each of the two contribution and the full fitting curve are plotted along with the data in Fig.~\ref{ris:fig2}(a), demonstrating an excellent quality of the fit. The peak position shows a clear shift when the sample temperature is varied. The peak position extracted from the fit is plotted in Fig.~\ref{ris:fig2}(b) as a function of $T$ in the whole range between 10~K and $T_{\text C}$ (222~K).

Interestingly, the propagation vector demonstrates a strictly linear change within the entire temperature range, starting at $q \approx 0.005$~\AA$^{-1}$~at high $T$ just below $T_{\text C}$ and shrinking by almost a factor of two to 0.003~\AA$^{-1}$~at $T = 10$~K. This corresponds to the period of the helical spin structure (and the skyrmion size) $\lambda = 2\pi/q$ varying from $\sim$120 to 210~nm [Fig~\ref{ris:fig2}(b)]. We examined the magnetic-texture periodicity in applied magnetic fields at different temperatures and also under different FC protocols and found that it depends only on the temperature and not on the field or field history. The Bragg peak intensity in the vicinity of $T_{\text C}$ is shown in Fig.~\ref{ris:fig2}(d). Since the intensity is proportional to $S^2$, it demonstrates a typical critical behavior of an order parameter, with the critical exponent $\beta = 0.33(2)$, in agreement with the theoretical value for the 3D Heisenberg model.

Having precisely determined the propagation vector in FePtMo$_3$N and its temperature variation, we turn to the field response of the magnetic texture. It was  shown in the previous study by magnetic measurements~\cite{Kautzsch} that the A phase, which is expected to host skyrmions, occurs upon an application of a magnetic field of $\sim$10--20~mT  in a temperature range of a few kelvin just below $T_{\text C}$. To explore if a metastable SkL state can also exist at lower temperatures and zero field, we measured SANS intensity with the following field protocol: (1) cool the sample in a field of $-13$~mT (through the stable SkL state) down to 214.5~K, (2) sweep the field from $-13$~mT to $+50$~mT, (3) sweep the field from $+50$~mT to $-50$~mT.

The results of these measurements are summarized in Figs.~\ref{ris:fig3}(a) and \ref{ris:fig3}(b) and the measurement protocol is schematically depicted on the determined phase diagram in Fig.~\ref{ris:fig3}(c). The critical fields $H_{C1}$, $H_A$, and $H_{C2}$ [Fig.~\ref{ris:fig3}(c)] denote the equilibrium (transitions in an applied field after ZFC) phase boundaries between the helical (H) and the conical (C) phases, C and SkL/SkL and C, and C and the polarized ferromagnetic (FM) state, respectively. Figure~\ref{ris:fig3}(a) shows a series of SANS patterns collected at different magnitudes of the applied field at 214.5~K after field cooling in a field of $-13$~mT. As can be seen by the intensity distribution, the magnetic system is driven  though an irreversible sequence of magnetic phases. At $B = -13$~mT, the pattern consists of two pairs of Bragg peaks. One pair of the peaks is oriented along the applied field direction, similarly to pattern shown in Fig~\ref{ris:fig1}(c), and implies a presence of the conical spiral phase. The other pair of the peaks are oriented perpendicular to the field and is a part of the ring of intensity formed in the reciprocal plane normal to the field. The latter is a hallmark of the SkL state~\cite{footnote,Sukhanov,Altynbaev2020}. It is worth mentioning that even though the conical intensity appears to be higher than the SkL intensity, the actual balance of the SkL/conical phase coexistence in the sample is significantly shifted towards the SkL phase. In other words, the skyrmions occupy the majority of the sample volume, since the SkL intensity is observed only at the intersection of the detector plane with the corresponding reciprocal space plane, whereas the whole intensity of the scattering from the minor conical phase is within the detector plane~\cite{Sukhanov}.

When the field is changed from $-13$~mT to 0~mT [Fig.~\ref{ris:fig3}(a)], the same pattern is recorded, which shows that the system remains in its initial state. The SkL intensity is almost fully transferred to the intensity of the conical state at $B = 11$~mT. This means that the metastable SkL decays when the field is reversed. Upon driving the system to the $B = 0$ condition, the helical state persists in the system, which is identical to the state at $T = 214.5$~K after the ZFC procedure. The field dependence of the conical and SkL intensities extracted from the SANS patterns is plotted along in Fig.~\ref{ris:fig3}(b). To plot the intensities of the conical and skyrmion Bragg peaks, we subsequently integrated the intensity of each peak over a square box that encloses the peak. As one can see, the SkL intensity stays unchanged between $B = -13$~mT (the initial state) and $B = 0$. This confirms that the compound exhibits zero-field skyrmions as a metastable state. The SkL state is preserved in a small magnitude of the reversed field ($\sim$2~mT) and transforms to the conical phase at $B \approx 8$~mT, then the fully-polarized state is reached at $\sim$30~mT. The transition between the metastable skyrmion phase and the conical phase on reversing the field after FC is labelled as $H_{A^{\prime}}$ on the phase diagram in Fig.~\ref{ris:fig3}(c).

Upon reversing the field, the conical intensity is closely reproduced until $\sim$14~mT, after which it starts deviating from the first run. When returning to zero field, the full helical intensity is recovered as the skyrmions are no longer nucleated. The intensities were further followed by sweeping the field to $-50$~mT. As one can see, only the conical-to-polarized transition is observed at the same critical field. The intensity profiles extracted from the SANS patterns of Fig.~\ref{ris:fig3}(a) are shown in Figs.~S1--S3 in supplemental material~\cite{supp} where clear positions of the conical and the SkL Bragg peaks can be determined.


To find out if the skyrmions in FePtMo$_3$N also exist in zero field at temperatures much lower than the lower-boundary temperature of the stable SkL phase ($\sim$5~K below $T_{\text C}$), we compared the SANS patterns at $T = 10$~K recorded after ZFC and FC protocols [Figs.~\ref{ris:fig4}(a)--\ref{ris:fig4}(d)]. In the FC protocol, the sample was cooled in $B = -13$~mT to 10~K, afterwards the field was driven to zero (see Fig.~\ref{ris:fig3}(c)). Clear Bragg peaks at momentum $\textbf{Q} \perp \textbf{B}$ are present in Fig.~\ref{ris:fig4}(a) confirming the presence of skyrmions, which is in contrast to the ZFC pattern in Fig.~\ref{ris:fig4}(b), where only the helical Bragg peaks are observed at $\textbf{Q} \parallel \textbf{B}$. To quantitatively compare the two patterns, the intensity profiles were plotted as a function of the azimuthal (in-plane) angle $\phi$ at fixed momenta $|\textbf{Q}| = q$ [Fig.~\ref{ris:fig4}(c)] and as a function of $Q$ at $\phi = 0^{\circ}$ and $180^{\circ}$ [Fig.~\ref{ris:fig4}(d)]. As can be seen in Fig.~\ref{ris:fig4}(c), additional peaks in the intensity profiles occur at $\phi$ that corresponds to the $\textbf{Q} \perp \textbf{B}$ relation. Both the $\phi$ and the $Q$ profiles of the SkL peaks are similar to those of the helical/conical-phase peaks.

The previous magnetic susceptibility study of the FePt$_{1-x}$Pd$_x$Mo$_3$N compounds allowed the estimation of the helical wavelength as a function of the concentration $x$~\cite{Kautzsch}. The estimation is possible in the framework of the Bak-Jensen model and the relation between the critical magnetic field $H_{{\text c}2}$ (the conical to field-polarized state transition) and the helical propagation vector $q$. As follows from this theory~\cite{Bak,Grigoriev15,Wilson,Maleyev}, the relation reads $g\mu_{\text B}H_{{\text c}2} = JSa^2q^2$, where $J$ is the effective exchange constant, $S$ is the ordered spin, and $a$ is the lattice constant. Because $J$ can be estimated from the Curie temperature~\cite{Kautzsch,Buhrandt}, one can calculate the predicted value of $q$ (and the spiral period $\lambda$)  from the experimental data on $H_{{\text c}2}$. Surprisingly, our measurements showed that the actual spin modulation period in FePtMo$_3$N is more than a factor of 3 longer at low temperature, $\lambda_{\text{exp}} \sim$210~nm, as compared to $\lambda_{\text{calc}} = 65$~nm. This is rather unusual for the cubic chiral magnets, as in many other compounds the Bak-Jensen equations were shown to give a good agreement~\cite{Kanazawa}.

The observed strictly linear temperature variation of the propagation vector is even more unusual. The period of the magnetic texture in the chiral magnets with $P2_13$ or $P4_132$ symmetry exhibits high susceptibility to chemical substitution~\cite{Bannenberg,Grigoriev14,Grigoriev13,Altynbaev,Grytsiuk,Karube,Karube2}, but shows either only slight or no temperature dependence. Therefore, FePtMo$_3$N represents a unique skyrmion host where the skyrmion size can be continuously tuned in a relatively broad range ($\lambda_{\text{max}} \approx 2\lambda_{\text{min}}$) by controlling the sample temperature.


To conclude, we showed that FePtMo$_3$N orders in a helical magnetic structure below $T_{\text C}$ of 222~K with the spiral period of $\sim$120~nm just below the $T_{\text C}$ and $\sim$210~nm at 5~K. Upon application of magnetic field at temperatures close to $T_{\text C}$, the helical structure of FePtMo$_3$N transforms into the SkL, confirming the interpretation of the recent magnetic susceptibility measurements~\cite{Kautzsch}. Moreover, the SkL shows a robust metastability down to low temperatures and zero field, placing FePtMo$_3$N next to the Co$_8$Zn$_8$Mn$_4$ and Co$_9$Zn$_9$Mn$_2$ compounds, whose properties were deemed unique up to date. The wide temperature-field range in which the metastable skyrmions are observed is possibly due to the site disorder of the Fe and Pt atoms in the crystal structure. 


\section*{Acknowledgments}

A.S.S. thanks S. E. Nikitin for the critical reading of the manuscript. This project was funded by the German Research Foundation (DFG) under Grants No. IN 209/9-1 and IN 209/7-1 as part of the Priority Program SPP 2137 ``Skyrmionics'', via the project C03 of the Collaborative Research Center SFB 1143 (project-id 247310070) at the TU Dresden and the W\"{u}rzburg-Dresden Cluster of Excellence on Complexity and Topology in Quantum Matter -- \textit{ct.qmat} (EXC 2147, project-id 39085490). The work at Santa Barbara was supported by the MRSEC Program of the National Science Foundation under Award No. DMR 1720256 (IRG-1). A.S.S. acknowledges support from the International Max Planck Research School for Chemistry and Physics of Quantum Materials (IMPRS-CPQM).

\end{document}